# Capacity Region of Layered Erasure One-sided Interference Channels without CSIT

Yan Zhu and Dongning Guo
Department of Electrical Engineering and Computer Science
Northwestern University, Evanston, IL 60208, USA
Email: {yan-zhu, dguo}@northwestern.edu

*Abstract*— This paper studies a layered erasure interference channel model, which is a simplification of the Gaussian interference channel with fading using the deterministic model approach. In particular, the capacity region of the layered erasure one-sided interference channel is completely determined, assuming that the channel state information (CSI) is known to the receivers, but there is no CSI at transmitters (CSIT). The result holds for arbitrary fading statistics. Previous results of Aggarwal, Sankar, Calderbank and Poor on the capacity region or sum capacity under several interference configurations are shown to be special cases of the capacity region shown in this paper.

## I. Introduction

Recent advances in the capacity of interference channels mainly focus on the situations where the channel state information (CSI) is known at both transmitters and receivers [1]–[3]. However, in practical system where the received signal suffers from fast fading, the CSI measured at the receivers can not be fed back to the transmitters accurately and in a timely manner. Furthermore, in two-way wireless systems using frequency division duplex (FDD) rather than time division duplex (TDD), one cannot take advantage of reciprocity of the wireless channel either.

In this paper, we consider an interference channel with two transmitter-receiver pairs, where the message of transmitter 1 is intended to receiver 1, and the message of transmitter 2 is intended to receiver 2. It is assumed that the interference is one-sided, so that transmission between transmitter 2 and receiver 2 is free of interference. Such a scenario may occur when, for example, receiver 1 is within the range of both transmitters, while receiver 2 is out of the range of transmitter 1.

As a precursor of an investigation of Gaussian interference channels with fading, we focus on the layered erasure model in this paper. This model is a variation of deterministic model introduced in [4] and can provide many insights on further study of the corresponding Gaussian model [5]. We note that such layered erasure channel with one-sided interference has been studied recently by Aggarwal *et al.* [6], where the capacity region or sum-capacity for some special cases is established. In particular, the authors of [6] have established the capacity region for uniformly very strong interference (Theorem 3 in [6]) and ergodic very strong interference (Theorem 6), and the sum-capacity for uniformly strong but not very strong interference (Theorem 4), uniformly weak interference (Theorem 7) and a special class of mixed interference (Theorem 9).

In this paper, the capacity region of the layered erasure one-sided interference channel is completely determined. To make the development easy to understand, the proof is first given for the special case of the single-layer erasure model, and then extended to the general layered erasure model. We also verify that several results in [6] are indeed special cases of the general result given in this paper.

## II. Model, Notation, and Main Results

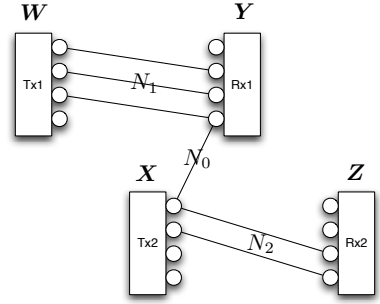

Fig. 1. Layered erasure channel with one-sided interference channel.

Consider a layered erasure channel model for the one-sided interference channel. Let the signals emitted by transmitters 1 and 2 at the $m$-th time interval be denoted by $\boldsymbol{W}[m]$ and $\boldsymbol{X}[m]$ respectively, which take values in $\mathbb{F}_2^q$. Let $\underline{s}$ denote a $q \times q$ matrix with $\underline{s}_{i+1,i} = 1$ for $i = 1, \ldots, q-1$ and all other elements being 0, so that $\underline{s}[x_1, x_2, \ldots, x_q]^\mathsf{T} = [0, x_1, \ldots, x_{q-1}]^\mathsf{T}$, and $\underline{s}^n \boldsymbol{X}[m]$ denotes a downward shift of the elements of the vector $\boldsymbol{X}[m]$ with $n$ least significant bits dropped out and $n$ zeros padded from the top of the vector. The received signals at time interval $m$ are then expressed as:

$$\boldsymbol{Y}[m] = \underline{s}^{q-N_1[m]} \boldsymbol{W}[m] \oplus \underline{s}^{q-N_0[m]} \boldsymbol{X}[m] \qquad (1a)$$
$$\boldsymbol{Z}[m] = \underline{s}^{q-N_2[m]} \boldsymbol{X}[m] \qquad (1b)$$

where $\{N_0[m]\}$, $\{N_1[m]\}$ and $\{N_2[m]\}$ are integer random processes taking values in $\{0, \ldots, q\}$, which represent the fading state of the three physical links. Let $(\{N_0[m]\}, \{N_2[m]\})$ and $\{N_1[m]\}$ be independent, and each of the three processes be independent and identically distributed (i.i.d.) over time (so that the channels are memoryless). It is further assumed that the fading states are known to both receivers but not to the transmitters.

For a random vector $\boldsymbol{X} \in \mathbb{F}_2^q$, let $X_i$ denote its $i$-th element and $\boldsymbol{X}_i^j$ denote $[X_i, \ldots, X_j]^\mathsf{T}$. For a vector process $\boldsymbol{X}[1], \ldots, \boldsymbol{X}[M]$, we use $(X_i)_l^k$ to denote the sequence $X_i[l], \ldots, X_i[k]$, and use $(\boldsymbol{X}_i^j)_l^k$ to denote the sequence $\boldsymbol{X}_i^j[l], \ldots, \boldsymbol{X}_i^j[k]$. The indexes outside the parentheses always refer to time. Binary addition of vectors of different length is aligned at the least significant bits; that is, if that $n_1 \geq n_2$, then define $\boldsymbol{X}_1^{n_1} \oplus \boldsymbol{W}_1^{n_2} = [X_1, \ldots, X_{n_1-n_2}, X_{n_1-n_2+1} \oplus W_1, \ldots, X_{n_1} \oplus W_{n_2}]^\mathsf{T}$.

*Theorem 1:* The capacity region of channel (1) is:

$$\mathcal{C} = \left\{ \begin{array}{c} 0 \leq R_1 \leq \mathbb{E}N_1 \\ 0 \leq R_2 \leq \mathbb{E}N_2 \\ (R_1, R_2): \; R_1 + \omega R_2 \leq \mathbb{E}N_1 + \omega \mathbb{E}\left[N_0 - N_1\right]^+ \\ + \sum_{l=1}^q \left(\omega \beta(l) - \alpha(l)\right)^+ \\ \forall \omega \in [0, 1] \end{array} \right\} \quad (2)$$

where, for every $l \in \{1, \ldots, q\}$,

$$\alpha(l) = \sum_{n_1 = 0}^q \mathsf{P}\left(N_1 = n_1\right) \mathsf{P}\left(l \leq N_0 < n_1 + l\right) \quad (3)$$

and

$$\beta(l) = \sum_{n_1 = 0}^q \mathsf{P}\left(N_1 = n_1\right)(\mathsf{P}\left(N_2 \geq l\right) - \mathsf{P}\left(N_0 \geq n_1 + l\right))^+. \quad (4)$$

In Section III, we proof the theorem for the single-layer case, *i.e.*, $q = 1$, to illustrate the key ideas. Due to space limitations, we only provide an outline of the proof for the general case in Section IV. The complete proof of Theorem 1 can be found in [7].

## III. THE SINGLE-LAYER ERASURE MODEL

Let $q = 1$. We denote the erasure probability of the link labeled by $N_i$ as $\epsilon_i$ and let $\bar{\epsilon}_i = 1 - \epsilon_i$ for notational convenience. Evidently $\bar{\epsilon}_i$ is the probability that the input symbol actually traverses the channel. Since $\boldsymbol{X} = X$ and $\boldsymbol{W} = W$ are scalars, and $N_i = 0$ or $1$, $i = 0, 1, 2$, we can denote $\boldsymbol{X}_1^{N_i}$ by $N_i X$ and $\boldsymbol{W}_1^{N_1}$ with $N_1 W$. By Theorem 1, we need to show $\mathcal{C}$ is the capacity region for $q = 1$. Specifically, $\alpha(1) = \bar{\epsilon}_0 \bar{\epsilon}_1$ and $\beta(1) = \bar{\epsilon}_1 \min(\bar{\epsilon}_2, \bar{\epsilon}_0) + (\bar{\epsilon}_2 - \bar{\epsilon}_0)^+$. If $\bar{\epsilon}_0 \geq \bar{\epsilon}_2$, $\mathcal{C}$ is the polyhedron with boundary constraints $0 \leq R_1 \leq \bar{\epsilon}_1$, $0 \leq R_2 \leq \bar{\epsilon}_2$, and

$$R_1 + R_2 \leq \bar{\epsilon}_0 + \bar{\epsilon}_1 - \bar{\epsilon}_0 \bar{\epsilon}_1. \quad (5)$$

If $\bar{\epsilon}_2 \geq \bar{\epsilon}_0$, $\mathcal{C}$ is the polyhedron with boundary constraints $0 \leq R_1 \leq \bar{\epsilon}_1$, $0 \leq R_2 \leq \bar{\epsilon}_2$, and

$$R_1 + \frac{\bar{\epsilon}_0 \bar{\epsilon}_1}{\beta(1)} R_2 \leq \bar{\epsilon}_1 + \frac{\bar{\epsilon}_0^2 \bar{\epsilon}_1 \epsilon_1}{\beta(1)}. \quad (6)$$

The capacity region is demonstrated in Fig. 2 for all possible scenarios depending on the parameters.

### A. Proof of Achievability

In each sub-figure of Fig. 2, we shadow the pentagon region enclosed by $R_1$-axis, $R_2$-axis, line $R_1 = \bar{\epsilon}_1$, line $R_2 = \bar{\epsilon}_0$, and line $R_1 + R_2 = 1 - \epsilon_0 \epsilon_1 = \bar{\epsilon}_0 + \bar{\epsilon}_1 - \bar{\epsilon}_0 \bar{\epsilon}_1$, which is the capacity region of the following multiple access channel (MAC):

$$Y = N_1 W \oplus N_0 X. \quad (7)$$

Note that if an achievable rate pair $(R_1, R_2)$ for channel (1) falls into the MAC capacity region, then the information of user 2 can be decoded at receiver 1. With these in mind, we investigate the achievability for all two possible cases:

If $\bar{\epsilon}_2 \leq \bar{\epsilon}_0$, $\mathcal{C}$ is contained in the MAC capacity region, (see Fig. 2(a) and Fig. 2(b)). Any rate pair in $\mathcal{C}$ can be achieved by using equiprobable inputs and allowing receiver 1 to decode information of both users.

If $\bar{\epsilon}_2 \geq \bar{\epsilon}_0$, it suffices to show that the two corner points $(\bar{\epsilon}_1, \bar{\epsilon}_0 \epsilon_1)$ and $(\epsilon_0 \bar{\epsilon}_1, \bar{\epsilon}_2)$, which are marked with star and square in Fig 2(c), respectively, are achievable. Because the first point is inside the MAC channel capacity region, it can be achieved. For the second point, we let both users use random codebook generated by binary equiprobable symbols. Let the code rate of user 2 be $\bar{\epsilon}_2$. Note that if $(N_0, N_1) = (0, 1)$, then $Y = W$; for all other realizations of $(N_0, N_1)$, $Y = W$ with probability $1/2$. Therefore, this is equivalent to an erasure channel with erasure probability $1 - \epsilon_0 \bar{\epsilon}_1$. Thus the rate $\epsilon_0 \bar{\epsilon}_1$ is achievable by user 1, which shows that the rate pair marked by the star can be achieved if $\bar{\epsilon}_2 \geq \bar{\epsilon}_0$.

### B. Proof of Converse

Any achievable rate pair $(R_1, R_2)$ must satisfy $R_1 \leq \bar{\epsilon}_1$ and $R_2 \leq \bar{\epsilon}_2$. For the strong-interference case where $\bar{\epsilon}_0 \geq \bar{\epsilon}_2$, it is sufficient to show that (5) holds and for weak-interference case where $\bar{\epsilon}_0 \leq \bar{\epsilon}_2$, it suffices to show that (6) holds.

It is easy to see that the capacity region of the one-sided interference channel only depends on the marginal conditional distribution of channel outputs at the receivers, but *not* on the joint distribution [8]. Without loss of generality, let the random variables $N_0$ and $N_2$ be "aligned" such that $\mathsf{P}(N_0 N_1 = 1) = \min(\bar{\epsilon}_0, \bar{\epsilon}_2)$. That is, if the realization of the weaker one between $N_0$ and $N_2$ is equal to 1, then the realization of the stronger one must also be equal to 1.

The converse of Theorem 1 for $q = 1$ is proved as follows. For notational simplicity, let $(\boldsymbol{N})^n$ denote all channel coefficients from time 1 to time $n$, *i.e.*, $(\boldsymbol{N})^n = \{N_i[j] : i = 0, 1, 2, \text{and } j = 1, \ldots, n\}$.

Consider first the case $\bar{\epsilon}_0 \geq \bar{\epsilon}_2$. By Fano's inequality,

$$\begin{aligned} nR_1 - n\delta_n &\leq \mathcal{I}\left(Y[1], \ldots, Y[n]; W[1], \ldots, W[n] | (\boldsymbol{N})^n\right) \\ &= \mathcal{H}\left((Y)_1^n | (\boldsymbol{N})^n\right) - \mathcal{H}\left((Y)_1^n | (W)_1^n, (\boldsymbol{N})^n\right) \\ &= \mathcal{H}\left((Y)_1^n | (\boldsymbol{N})^n\right) - \mathcal{H}\left((N_0 X)_1^n | (\boldsymbol{N})^n\right) \\ &\leq n(\bar{\epsilon}_1 + \bar{\epsilon}_0 - \bar{\epsilon}_0 \bar{\epsilon}_1) - \mathcal{H}\left((N_0 X)_1^n | (\boldsymbol{N})^n\right) \quad (8) \end{aligned}$$

where $\delta_n$ vanishes as $n \to \infty$. The last inequality follows from that $\mathcal{H}\left((Y)_1^n | (\boldsymbol{N})^n\right)$ is maximized by setting both $(W)_1^n$ and $(X)_1^n$ to be i.i.d Bernoulli $\left(\frac{1}{2}\right)$ sequence.

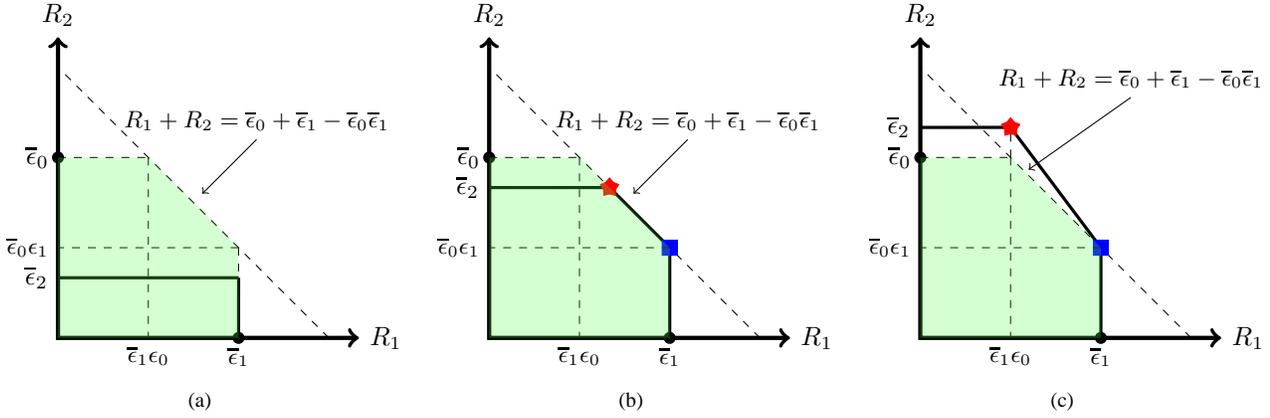

Fig. 2. Capacity region for single-layer erasure channel with different cases drawn by solid lines. (a) $\bar{\epsilon}_2 \leq \bar{\epsilon}_0 \epsilon_1$. (b) $\bar{\epsilon}_0 \geq \bar{\epsilon}_2 \geq \bar{\epsilon}_0 \epsilon_1$. (c) $\bar{\epsilon}_2 \geq \bar{\epsilon}_0$.

Also due to Fano's inequality, we have

$$nR_2 - n\delta_n \leq \mathcal{I}((Z)_1^n; (X)_1^n | (\mathbf{N})^n)$$
$$= \mathcal{H}((N_2 X)_1^n | (\mathbf{N})^n). \quad (9)$$

Since $\bar{\epsilon}_0 \geq \bar{\epsilon}_2$ by assumption, we have $N_0 \geq N_2$ and thus $\mathcal{H}((N_0 X)_1^n | (\mathbf{N})^n) \geq \mathcal{H}((N_2 X)_1^n | (\mathbf{N})^n)$. Comparing (8) and (9) yields

$$nR_1 + nR_2 - 2n\delta_n \leq n(\bar{\epsilon}_1 + \bar{\epsilon}_0 - \bar{\epsilon}_0 \bar{\epsilon}_1)$$

which completes the proof of (5) by noting that $\delta_n \to 0$ as $n \to \infty$.

Consider next the case of $\bar{\epsilon}_2 \geq \bar{\epsilon}_0$. Let $(\widetilde{W})_1^n = (\widetilde{W}[1], \ldots, \widetilde{W}[n])$ be i.i.d. Bernoulli $(\frac{1}{2})$ sequence independent of $(X)_1^n$ and let $\widetilde{Y} = N_1 \widetilde{W} \oplus N_0 X$. Applying Fano's inequality, we obtain

$$nR_1 - n\delta_n$$
$$\leq \mathcal{I}((Y)_1^n; (W)_1^n | (\mathbf{N})^n)$$
$$= \mathcal{H}((N_1 W \oplus N_0 X)_1^n | (\mathbf{N})^n) - \mathcal{H}((N_0 X)_1^n | (\mathbf{N})^n)$$
$$\leq \mathcal{H}\left((N_1 \widetilde{W} \oplus N_1 W \oplus N_0 X)_1^n \big| (\mathbf{N})^n\right) - \mathcal{H}((N_0 X)_1^n | (\mathbf{N})^n)$$
$$= \mathcal{H}\left((N_1 \widetilde{W} \oplus N_0 X)_1^n \big| (\mathbf{N})^n\right) - \mathcal{H}((N_0 \mathbf{X})_1^n | (\mathbf{N})^n)$$
$$= \mathcal{I}\left((\widetilde{Y})_1^n; (\widetilde{W})_1^n \big| (\mathbf{N})^n\right)$$

where the second inequality follows from data processing theorem and the second equality is due to the fact that $\widetilde{W} \oplus W$ is identically distributed as $\widetilde{W}$. Since $(\widetilde{W})_1^n$—$(N_1 \widetilde{W})_1^n$—$(\widetilde{Y})_1^n$ is a Markov chain, due to data processing theorem and Fano's inequality,

$$nR_1 - n\delta_n$$
$$\leq \mathcal{I}\left((\widetilde{Y})_1^n; (N_1 \widetilde{W})_1^n \big| (\mathbf{N})^n\right)$$
$$= \mathcal{H}\left((N_1 \widetilde{W})_1^n \big| (\mathbf{N})^n\right) - \mathcal{H}\left((N_1 \widetilde{W})_1^n \big| (\widetilde{Y})_1^n, (\mathbf{N})^n\right)$$
$$\leq n\bar{\epsilon}_1 - \mathcal{H}\left((N_1 \widetilde{W})_1^n \big| (\widetilde{Y})_1^n, (\mathbf{N})^n\right)$$
$$= n\bar{\epsilon}_1 - \mathcal{H}\left((N_0 X)_1^n \big| (\widetilde{Y})_1^n, (\mathbf{N})^n\right). \quad (10)$$

Also by Fano's inequality,

$$nR_2 - n\delta_n$$
$$\leq \mathcal{I}((N_2 X)_1^n; (X)_1^n | (\mathbf{N})^n)$$
$$\leq \mathcal{I}\left((N_2 X)_1^n, (\widetilde{Y})_1^n; (X)_1^n \big| (\mathbf{N})^n\right)$$
$$= \mathcal{I}\left((\widetilde{Y})_1^n; (X)_1^n \big| (\mathbf{N})^n\right) + \mathcal{I}\left((N_2 X)_1^n; (X)_1^n \big| (\widetilde{Y})_1^n, (\mathbf{N})^n\right)$$
$$\leq n\bar{\epsilon}_0 \epsilon_1 + \mathcal{H}\left((N_2 X)_1^n \big| (\widetilde{Y})_1^n, (\mathbf{N})^n\right). \quad (11)$$

The upper bounds (10) and (11) can be understood as follows: Comparing with the interference free scenario, the rate loss of user 1 due to the interference signal $X$ is at least $\mathcal{H}\left((N_0 X)_1^n \big| (\widetilde{Y})_1^n, (\mathbf{N})^n\right)$. The maximum rate of user 2 is $\bar{\epsilon}_0 \epsilon_1$ if the signal $X$ is decodable by user 1. Since the signal $X$ is not required to be decodable by user 1, the additional rate encoded in $X$ for user 2 is $\mathcal{H}\left((N_2 X)_1^n \big| (\widetilde{Y})_1^n, (\mathbf{N})^n\right)$. In the following, we consider the trade-off between the rate loss of user 1 and the rate gain of user 2 over all choices of the signal $(X)_1^n$. Using Marton-like expansion [9], [10]

$$\mathcal{H}\left((N_2 X)_1^n \big| (\widetilde{Y})_1^n, (\mathbf{N})^n\right) - \mathcal{H}\left((N_0 X)_1^n \big| (\widetilde{Y})_1^n, (\mathbf{N})^n\right)$$
$$= \sum_{i=1}^{n} \left\{ \mathcal{H}\left((N_2 X)_1^i, (N_0 X)_{i+1}^n \big| (\widetilde{Y})_1^n, (\mathbf{N})^n\right) \right.$$
$$\left. - \mathcal{H}\left((N_2 X)_1^{i-1}, (N_0 X)_i^n \big| (\widetilde{Y})_1^n, (\mathbf{N})^n\right) \right\}.$$

Moreover, using the chain rule,

$$\mathcal{H}\left((N_2 X)_1^n \big| (\widetilde{Y})_1^n, (\mathbf{N})^n\right) - \mathcal{H}\left((N_0 X)_1^n \big| (\widetilde{Y})_1^n, (\mathbf{N})^n\right)$$
$$= \sum_{i=1}^{n} \left\{ \mathcal{H}\left((N_2 X)_i \big| (N_2 X)_1^{i-1}, (N_0 X)_{i+1}^n, (\widetilde{Y})_1^n, (\mathbf{N})^n\right) \right.$$
$$\left. - \mathcal{H}\left((N_0 X)_i \big| (N_2 X)_1^{i-1}, (N_0 X)_{i+1}^n, (\widetilde{Y})_1^n, (\mathbf{N})^n\right) \right\}. \quad (12)$$

To bound (12), we need following lemma:

*Lemma 1:* Let $T$ be a collection of random variables which are independent of $(N_0, N_1, N_2)$. Let $\widetilde{W}$ be a Bernoulli $(\frac{1}{2})$

r.v. independent of $X$. Then
$$\mathcal{H}\left(N_0 X | N_0 X \oplus N_1 \widetilde{W}, T, \boldsymbol{N}\right) = \bar{\epsilon}_0 \bar{\epsilon}_1 \mathcal{H}(X|T)$$
and
$$\mathcal{H}\left(N_2 X | N_0 X \oplus N_1 \widetilde{W}, T, \boldsymbol{N}\right) = \beta(1) \mathcal{H}(X|T).$$
Therefore,
$$\mathcal{H}\left(N_0 X | N_0 X \oplus N_1 \widetilde{W}, T, \boldsymbol{N}\right) = \frac{\bar{\epsilon}_0 \bar{\epsilon}_1}{\beta(1)} \mathcal{H}\left(N_2 X | N_0 X \oplus N_1 \widetilde{W}, T, \boldsymbol{N}\right).$$

The proof is straightforward and omitted here. Note that if $\bar{\epsilon}_2 \geq \bar{\epsilon}_0$, $\beta(1) = \bar{\epsilon}_2 - \bar{\epsilon}_0 \epsilon_1 \geq \bar{\epsilon}_0 \bar{\epsilon}_1$.

For each $i$, applying Lemma 1 to (12) with
$$T_i = \left((N_2 X)_1^{i-1}, (N_0 X)_{i+1}^n, (\widetilde{Y})_1^{i-1}, (\widetilde{Y})_{i+1}^n, (\boldsymbol{N})_1^{i-1}, (\boldsymbol{N})_{i+1}^n\right)$$
which is independent of $(N_0)_i$, $(N_1)_i$, and $(N_2)_i$, we obtain
$$\mathcal{H}\left((N_2 X)_1^n \middle| (\widetilde{Y})_1^n, (\boldsymbol{N})^n\right) - \mathcal{H}\left((N_0 X)_1^n \middle| (\widetilde{Y})_1^n, (\boldsymbol{N})^n\right)$$
$$= \sum_{i=1}^n \left(1 - \frac{\bar{\epsilon}_0 \bar{\epsilon}_1}{\beta(1)}\right) \mathcal{H}\left((N_2 X)_i \middle| (\widetilde{Y})_i, T_i, \boldsymbol{N}_i\right)$$
$$\leq \sum_{i=1}^n \left(1 - \frac{\bar{\epsilon}_0 \bar{\epsilon}_1}{\beta(1)}\right) \mathcal{H}\left((N_2 X)_i \middle| (N_2 X)_1^{i-1}, (\widetilde{Y})_1^n, (\boldsymbol{N})^n\right)$$
$$= \left(1 - \frac{\bar{\epsilon}_0 \bar{\epsilon}_1}{\beta(1)}\right) \mathcal{H}\left((N_2 X)_1^n \middle| (\widetilde{Y})_1^n, (\boldsymbol{N})^n\right)$$
where the last inequality is due to the fact that conditioning reduces entropy and the last equality is due to the chain rule. Thus, we have established
$$\mathcal{H}\left((N_0 X)_1^n \middle| (\widetilde{Y})_1^n, (\boldsymbol{N})^n\right) \geq \frac{\bar{\epsilon}_0 \bar{\epsilon}_1}{\beta(1)} \mathcal{H}\left((N_2 X)_1^n \middle| (\widetilde{Y})_1^n, (\boldsymbol{N})^n\right). \tag{13}$$

Comparing (10), (11), and (13), we have
$$n R_1 + \frac{\bar{\epsilon}_1 \bar{\epsilon}_0}{\beta(1)} n R_2 - n \delta_n - \frac{\bar{\epsilon}_1 \bar{\epsilon}_0}{\beta(1)} n \delta_n \leq \bar{\epsilon}_1 + \frac{\bar{\epsilon}_0^2 \bar{\epsilon}_1 (1 - \bar{\epsilon}_1)}{\beta(1)}.$$

As $n \to \infty$, we complete the proof of (6).

## IV. THE GENERAL LAYERED ERASURE MODEL

### A. Converse

With the insights from single-layer case, we first need to align the fading states. For any random variable $N$, let $F_N(n) = \mathsf{P}(N \leq n)$ and $\overline{F}_N(n) = \mathsf{P}(N \geq n)$. Also let $F_N^{-1}(t) = \inf\{u : F_N(u) \geq t\}$. Let $\Lambda$ be a r.v. uniformly distributed on interval $[0, 1]$, then we know that $\widetilde{N} := F_N^{-1}(\Lambda)$ is identically distributed with $N$. Because the capacity region depends only on the marginal distributions of the two received signals, we can assume the following alignment without changing the capacity region:
$$N_0 = F_{N_0}^{-1}(\Lambda), \quad \text{and} \quad N_2 = F_{N_2}^{-1}(\Lambda).$$

With the fading states aligned, we need following lemma for general $q$, which is parallel to Lemma 1.

*Lemma 2:* Let $T$ be a random variable such that it is independent of $N_0$, $N_1$ and $N_2$. Let $(\widetilde{\boldsymbol{W}})_1^n$ as an i.i.d. random sequence. For each $i$, $(\widetilde{\boldsymbol{W}})_i \in \mathbb{F}_2^q$ and elements of $(\widetilde{\boldsymbol{W}})_i$ are i.i.d. Bernoulli $\left(\frac{1}{2}\right)$ random variables. Furthermore, $\boldsymbol{X}$ be a random vector in $\mathbb{F}_2^q$ independent of $\widetilde{\boldsymbol{W}}$. Then
$$\mathcal{H}\left(\boldsymbol{X}_1^{N_0} \middle| \boldsymbol{X}_1^{N_0} \oplus \widetilde{\boldsymbol{W}}_1^{N_1}, T, N\right) = \sum_{l=1}^q \alpha(l) \mathcal{H}\left(X_l \middle| \boldsymbol{X}_1^{l-1}, T\right)$$
$$\mathcal{H}\left(\boldsymbol{X}_1^{N_2} \middle| \boldsymbol{X}_1^{N_0} \oplus \widetilde{\boldsymbol{W}}_1^{N_1}, T, N\right) = \sum_{l=1}^q \beta(l) \mathcal{H}\left(X_l \middle| \boldsymbol{X}_1^{l-1}, T\right)$$
where, for every $l \in \{1, \ldots, q\}$, $\alpha(l)$ and $\beta(l)$ are given in (3) and (4), respectively.

The proof is through straightforward calculation. For details, see [7].

By using Marton-like expansion and applying Lemma 2, we can show the converse of Theorem 1. The detailed proof is omitted here.

### B. Achievability

We next investigate the structure of the capacity region and briefly give the capacity-achievable coding schemes.

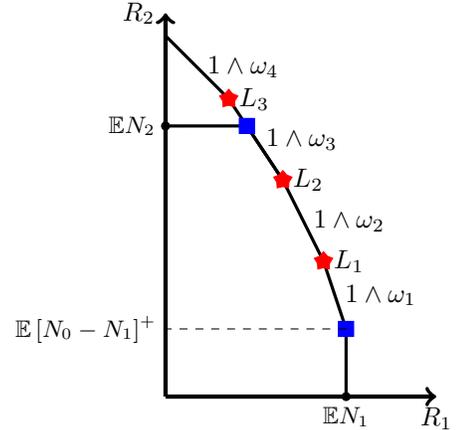

Fig. 3. An example of the capacity region, where $x \wedge y = \min(x, y)$.

*1) Structure of the Capacity Region:* An example of the capacity region is shown in Fig. 3. Generally, it is enclosed by three curves: two of them are lines $R_1 = \mathbb{E} N_1$ and $R_2 = \mathbb{E} N_2$; the third one is a piece-wise linear curve (denoted by $\mathcal{L}$), which can be obtained through the third constraint in (2) as follows. For fixed $R_1$, we find the maximum rate $R_2$ that can be supported.

For a closer examination of the curve $\mathcal{L}$, we define $\omega(l) = \alpha(l)/\beta(l)$ and order $\{\omega(l)\}_1^q$ as $\omega_1 \leq \ldots \omega_b < 1 \leq \omega_{b+1} \cdots \leq \omega_q$. Then curve $\mathcal{L}$ can be divided into $b + 1$ parts[1], each of which is a segment or a ray with slope $1/\omega_k$ or 1, where $k = 1, \ldots, b$. Since each of these segments or rays is on a line determined by its slope, we simply name these segments or rays with $\omega_k$. Let $L_i$ be the intersection point of line $\omega_i$ and line $\omega_{i+1}$. ($L_b$ is the intersection point of line $\omega_1$ and the line with slope 1. ) It is easy to see that $\{L_i\}_1^b$ are all extreme points on curve $\mathcal{L}$. And they are marked by stars in Fig. 3.

---
[1]There are some degenerate cases where some $\omega_k$ take the same value. Therefore, there can be fewer.

We observe that the curve $\mathcal{L}$ always intersects with line $R_1 = \mathbb{E}N_1$ at point $(\mathbb{E}N_1, \mathbb{E}[N_0 - N_1]^+)$. We refer to it as the *starting point* which is marked by a square in Fig 3. On the other hand, line $R_2 = \mathbb{E}N_2$ may intersect with the curve $\mathcal{L}$ or line $R_1 = \mathbb{E}N_1$ in various position, which may make some constraints with redundant, *e.g.*, in Fig. 3 constraint with $1 \wedge \omega_4$ is redundant. We call the intersection of $R_2 = \mathbb{E}N_2$ and curve $\mathcal{L}$ (or line $R_1 = \mathbb{E}N_1$) the *end point* and also mark it with square in Fig 3. Since we can exploit time sharing, it suffices to show the achievability for starting point, end point and extreme points in-between.

*2) Achievable Schemes:* The achievable scheme is basically the Han-Kobayashi scheme [11]. We split the message of transmitter 2 into a private message and a common message.

To achieve the starting point, it suffices to let user 2 transmit common message only.

To achieve the mid-way point $L_k$, let $\tau$ be a permutation on $\{1, \ldots, q\}$ such that $\omega(\tau(k)) = \omega_k$. To achieve point $L_k$, let user 2 transmit the private message on layers $\mathcal{B} := \{\tau(1), \ldots \tau(k)\}$ and transmit the common message on layers in $\mathcal{B}^c$.

The following considerations achieve the end point:
- If the end point is on line $R_1 = \mathbb{E}N_1$, let user 2 transmit the common message only.
- If end point is on a segment of curve $\mathcal{L}$ with slope steeper than 1, define $\mathcal{B} := \{\tau(1), \ldots \tau(k-1)\}$ and $\mathcal{U} := \{\tau(k+1), \ldots, \tau(q)\}$. Layers in $\mathcal{B}$ and $\mathcal{U}$ are used for the private message and the common message, respectively. For layer $\tau(k)$, we split it further into two parts: one for the private message and the other for the common message.
- If the end point is on the ray with slope 1, we let user 2 transmit the private message on layers $\mathcal{B} := \{\tau(1), \ldots \tau(b)\}$ and transmit the common message on the remaining layers. Different from other cases, we need to split user 1's information into two virtual users [12].

*C. Examples*

Before end of this section, we investigate some special cases for the layered erasure channel with one-sided interference.

*1) The Case of Stochastically Strong Interference:* We assume that $N_0 \geq_{st} N_2$, *i.e.*, $\overline{F}_{N_0}(l) \geq \overline{F}_{N_2}(l)$ for all $l \in \{1, \ldots, q\}$.

Evaluating the single-user constraints gives $R_i \leq \mathbb{E}N_i$, $i = 1, 2$. For the third constraint in (2), since $N_0 \geq_{st} N_2$, $\omega\beta(l) - \alpha(l) \leq 0$ for all $l \in \{1, \ldots, q\}$ and $\omega \in [0, 1]$. It easy to see that for every $\omega \in [0, 1]$, the line

$$R_1 + \omega R_2 = \mathbb{E}N_1 + \omega\mathbb{E}(N_0 - N_1)^+$$

always passes through the point $(\mathbb{E}N_1, \mathbb{E}(N_0 - N_1)^+)$. Therefore, the third bound in (2) can be reduced to the same bound with $\omega = 1$. Hence, the capacity region for stochastic strong interference can be simplified to

$$0 \leq R_i \leq \mathbb{E}N_i \quad i = 1, 2$$
$$R_1 + R_2 \leq \mathbb{E}\max(N_0, N_1)$$

which is a generalization of Theorems 3-5 in [6].

*2) The Case of Stochastically Weak Interference:* We assume that $N_0 \leq_{st} N_2$, *i.e.*, $\overline{F}_{N_0}(l) \leq \overline{F}_{N_2}(l)$ for all $l \in \{1, \ldots, q\}$. Therefore, $\beta(l) \geq \alpha(l)$ for all $l \in \{1, \ldots, q\}$. From the analysis on geometric structure of capacity region, the capacity region can be represented by

$$0 \leq R_i \leq \mathbb{E}N_i \quad i = 1, 2$$
$$R_1 + \omega_k R_2 \leq \mathbb{E}N_1 + \omega_k \mathbb{E}(N_0 - N_1)^+ + \sum_{l=1}^{q}(\omega_k \beta(l) - \alpha(l))^+$$

where $k \in \{1, \ldots, q\}$. Furthermore, the sum-capacity is achieved at the end point. Therefore,

$$C_{sum} = \mathbb{E}N_1 + \mathbb{E}(N_0 - N_1)^+ + \sum_{l=1}^{q}(\beta(l) - \alpha(l))$$
$$= \mathbb{E}\max(N_0, N_1) + \mathbb{E}N_2 - \mathbb{E}N_0$$

which is a generalization of Theorem 7 in [6].

V. CONCLUSION

In this work, we fully characterize the capacity region of the layered erasure one-sided interference channel with fading states known at receivers only. The converse is inspired by the Marton-like expansion while the achievability is based on HK schemes. The subsequent work [7] will show that the insights obtained here can be used to establish a constant gap capacity result for the corresponding Gaussian fading channel.